\documentclass[aps,preprint,prd]{revtex4}
\usepackage{graphicx}

\newcommand{\Lie}{{\cal L}}
\newcommand{\RR}{{\cal R}}
\newcommand{\Div}{\textrm{div }}
\newcommand{\RRmin}{{\cal R}_{min}}
\newcommand{\const}{\textrm{const}}
\newcommand{\dd}{\textrm{d}}

\begin{document}


\title{Isolated and dynamical horizons from a common perspective}


\author{Miko\l{}aj Korzy\'nski}
\affiliation{Institute of Theoretical Physics, Warsaw University}

\email{mkorz@fuw.edu.pl}

\date{\today}

\begin{abstract}
A framework is developed in which one can write down the constraint equations
on a three--dimensional hypersurface of arbitrary signature. It is  
then applied to isolated and dynamical horizons. The derived equations
can be used to extract physicaly relevant quantities describing the horizon
irrespective to whether it is isolated (null) or dynamical at a given instant of time.
Furthermore, small perturbation of isolated horizons are considered, and
finally a family of axially--symmetric exact solution of the constraint equations
on a dynamical horizon is presented. 

\end{abstract}


\maketitle

\section{Introduction}
\label{secIntroduction}

Since their introduction, isolated and dynamical horizons have been used to describe
black holes in general, dynamical situations, in both analytic \cite{ashtekar-2004-7,ashtekar-2000-85,ashtekar-2001-64,ashtekar-2004-21,korzynski-2005-22} and numerical \cite{jaramillo-2004-70,schnetter-2006-,gourgoulhon-2006-423} context. Both are 
defined as three--dimensional hypersurfaces foliated by mariginally trapped surfaces (so
called mariginally trapped tubes, MTT's).
They are supposed to represent the boundary of a black hole in a non--stationary spacetime. 
Isolated horizons (IH) are null hypersurfaces and describe  black holes which are in equilibrium
with the environment at the moment. Dynamical horizons (DH) on the other hand
are spacelike and describe a black hole absorbing matter or gravitational radiation.

In physical situations we expect an isolated horizon to become dynamical if matter or
gravitational radiation falls on it. On the other hand, it is believed that a perturbed
black hole will return to a stationary state after the matter inflow ends and all gravitational radiation is radiated away. In this case a DH should asymptotically become null. In both processes we are led to investigate a hypersurface which can change its
signature (from null to spacelike) depending on circumstances.

However, geometrically null and spacelike surfaces are very different. The latter are
equipped with a well--defined, positive definite metric, which makes them 
Riemannian manifolds. Such hypersurfaces are completely described
by their metrics and extrinsic curvature. On the other hand, on null surfaces one
usually uses a different set of geometrical objects \cite{ashtekar-2004-7}.

It is tempting to use perturbative techniques to investigate small perturbations of an IH
\cite{booth-2004-92, kavanagh-2006-,poisson-2004-70, poisson-2005-94}. This is not easy, as in presence of smallest
perturbations the IH is not strictly null anymore.
On the other hand, the ortogonal
vector cannot be normalized if a DH becomes null and therefore the standard definitions of
extrinsic curvature and covariant deriavative do not make sense any more \cite{gourgoulhon-2006-423}.

This demonstrates the need to introduce a way of description of a MTT (or any hypersurface) which is applicable to any signature. It should involve ``signature--blind''
variables which do not refer to any normalization of its ortogonal vector or to
the three-dimensional metric on the horizon. It should however contain all information needed
to recover the standard geometrical objects if the surface happens to be spacelike.

Although variables of that kind have been used in calculations by several authors 
\cite{booth-2004-92,booth-2005-22,gourgoulhon-2005-72}, there hasn't been
a complete discussion of the Einstein constraint equations in this context, 
neither on MTT's, nor in general. The aim of this paper is to 
 introduce such variables and discuss the Einstein constraint equations on
any hypersurface -- spacelike, null or timelike. We then focus on MTT's. We also analyse a lineary perturbed IH and finally present
a family of axialy symmeric, exact solutions of the Einstein constraint equations. 

\section{``Signature--blind'' variables}
\label{secVariables}

We consider a 3--dimensional hypersurface $H$ foliated by a family of 2--dimensional spacelike 
surfaces of spherical topology, called sections. The induced metric on $H$ may have arbitrary signature, or even be degenerate. 

By convention, the greek indeces will denote spacetime objects, while the latin indeces
objects defined on sections. In particular, if $X_\alpha$ is a one--form defined on 
spacetime, $X_A$ will denote its pullback to a section, whereas if $X^\alpha$ is a spacetime vector, $X^A$ will be the ortogonal projection to the tangent space to sections.

On each section we have a positive definite metric $q_{AB}$. 
At each point we can pick up two null vectors, $k^\mu$ and $l^\mu$, ortogonal to the leaves of the foliation and normalized $l_\mu\,k^\mu=-1$. We also choose a vector field $n^\alpha$ tangent to $H$, preserving the foliation and ortogonal to sections.

We may now decompose $n$ according to $n^\mu = n_l\, l^\mu + n_k\, k^\mu$.
The vector field $\tau^\mu = n_l\, l^\mu - n_k \ k^\mu$ is ortogonal to both leaves of the
foliation and $n$, and therefore ortogonal to $H$.

The choice of $l$ and $k$ is not unique, as we may rescale them simultanously by any
non--vanishing function $A$:
$l'^{\mu} = A \, l^{\mu}$, $k'^{\mu} = A^{-1} \, k^{\mu}$. This gauge freedom can be partially fixed by 
first fixing $l$ and $k$ at one section and then imposing the conditions
\begin{eqnarray}
\nabla_n\,l^A &=& - \nabla^A l_\mu \, k^\mu \nonumber\\
\nabla_n\,l^\alpha\,l_\alpha &=& \nabla_n\,l^\alpha\,k_\alpha = 0 \nonumber\\
\nabla_n\,k^A &=& - \nabla^A k_\mu \, l^\mu \nonumber\\
\nabla_n\,k^\alpha\,l_\alpha &=& \nabla_n\,k^\alpha\,k_\alpha = 0 \label{eqGaugeFixing}.
\end{eqnarray}

They can be checked to preserve the normalization of $l$ and $k$, their null character and  
their ortogonality to the foliation (see Appendix \ref{appGauge}).

We have therefore specified the $n$ component of the covariant derivative of $l$ and $k$.
Its pullback to sections constitutes so called Weingarten tensors or deformation tensors
\begin{eqnarray}
\nabla_A l_B &=& L_{AB} =  \sigma^l_{(AB)} + \frac{1}{2}\theta^l \, q_{AB} \\
\nabla_A k_B &=& K_{AB} =  \sigma^k_{(AB)} + \frac{1}{2}\theta^k \, q_{AB} . 
\end{eqnarray}
They have been decomposed to traceless, symmetric part, called shear, and the trace
called expansion. The antisymmetric part vanished since $k$ and $l$ are surface--forming.

We also have the rotation one--form $\omega_A = -\nabla_A l^\mu\,k_\mu =
\nabla_A k^\mu\,l_\mu$.

If $H$ is spacelike, the data specified here are equivalent to the standard objects
in constraint equations, \emph{i.e.} the three--dimensional metric and the extrinsic curvature. 
The precise transformation is given in Appendix \ref{appK}.

We still have the gauge freedom to multiply $l^\mu$ and $k^\mu$ by one function \emph{at a given
section}:
\begin{eqnarray}
l'^\mu &=& F\, l^\mu \nonumber\\
k'^\mu &=& F^{-1}\, k^\mu \nonumber\\
\omega'_A &=& \omega_A - D_A \ln |F| \label{eqGauge}
\end{eqnarray}
 
\section{The Einstein constraint equations}
\label{secConstraints}

The constraint equations on any three--dimensional hypersurface are the Einstein equations involving the
$G_{\alpha\beta} \tau^\alpha$ component of the Einstein tensor, where $\tau$ is ortogonal to the surface.

The $G_{\alpha\beta} \tau^\alpha$ one--form can be decomposed to the pullback to sections
$G_{\alpha B}\,\tau^\alpha$ and two sections--ortogonal components $G_{\alpha\beta} \, \tau^\alpha l^\beta$
and $G_{\alpha \beta} \tau^\alpha k^\beta$.

We shall denote the covartiant derivative on the section 2--spheres, compatible with $q_{AB}$, by 
$D$, its Laplacian $q^{AB} \, D_A\,D_B$ by $\Delta$ and its Ricci scalar by $\RR$.
The constraint equations read 
\begin{eqnarray}
\Lie_n \,\omega_A &=& G_{\mu A} \tau^\mu - n_l(D_B\,\bar L^B\!_A+\omega_A\, \theta^l)
 - D_B n_l\, L^B\!_A + \nonumber\\
 && + n_k (D_B\,\bar K^B\,_A - \omega_A\, \theta^k) + D_B n_k\, K^B\!_A \label{eqOmega0}\\
\Lie_n \theta^l &=& -G_{\mu\nu} \tau^\mu l^\nu + \Delta n_k - 2\omega^A \, D_A n_k
-n_k \left( \frac{\cal R}{2} + D_A\,\omega^A - \omega_A\,\omega^A + \theta^l
\,\theta^k\right)+\nonumber\\
&&-n_l\, L_{AB}\,L^{AB}\label{eqThetaL0}\\ 
\Lie_n \theta^k &=& G_{\mu\nu} \tau^\mu k^\nu + \Delta n_l + 2\omega^A \, D_A n_l
-n_l \left( \frac{\cal R}{2} - D_A\,\omega^A - \omega_A\,\omega^A + \theta^l
\,\theta^k\right)+\nonumber\\
&&-n_k\, K_{AB}\,K^{AB}\label{eqThetaK0}.
\end{eqnarray}
We have introduced the notation $\bar K_{AB} = \sigma^k_{AB} - \frac{1}{2}\theta^k\,q_{AB}$
and analogically for $\bar L_{AB}$.
These equations, together with the evolution equation for the metric $q_{AB}$
\begin{equation}
\Lie_{n} \,q_{AB} = 2n_l\, L_{AB} + 2n_k\, K_{AB} \label{eqQ0}
\end{equation}
constitute a system, which yields initial data for Cauchy problem in the spacelike case. It can be analysed and solved, as we shall see, at least in some special cases.

If we introduce a ``time'' coordinate on $H$ compatible with $n$ and the foliation, 
\emph{i.e.} $t$ is constant on sections and $n^\mu t,_\mu = 1$, we can replace the Lie derivatives in
(\ref{eqOmega0})-(\ref{eqQ0}) by ordinary partial derivatives with respect to $t$.

Equation (\ref{eqOmega0}) was discussed in \cite{gourgoulhon-2006-423} in
the null case and in \cite{gourgoulhon-2005-72} in a general situation.
Equations (\ref{eqThetaL0}) and (\ref{eqThetaK0}) appeared in \cite{eardley-1998-57} and
\cite{kavanagh-2006-}.

\section{Constraint equations on a MTT}
\label{secApparentH}

We now set $\theta_l = 0$. The equations (\ref{eqOmega0})--(\ref{eqQ0}) become now
\begin{eqnarray}
\Lie_n q_{AB} &=& 2n_l \,\sigma_{AB} + 2n_k \, K_{AB} \label{eqQ}\\
\Lie_n \,\omega_A &=& G_{\mu A} \tau^\mu - D_B(n_l\,\sigma^l\,^B\!_A) + n_k (D_B\,\bar K^B\,_A - \omega_A\, \theta^k) + D_B n_k\, K^B\!_A \label{eqOmega}\\
0 &=& -G_{\mu\nu} \tau^\mu l^\nu + P\,n_k - n_l \, \sigma^l_{AB}\,\sigma^l\,^{AB}\label{eqThetaL}\\ 
\Lie_n \theta^k &=& G_{\mu\nu} \tau^\mu k^\nu + P^*\, n_l-n_k\,
\left( \sigma^k_{AB}\,\sigma^k\,^{AB} + \frac{1}{2} \theta^k\,^2\right)\label{eqThetaK}.
\end{eqnarray}
We have introduced two elliptic differential operators
\begin{eqnarray}
P &=& \Delta - 2 \omega^A \, D_A - \left(\frac{\RR}{2} + \Div \omega - \omega^2\right)\\
P^* &=& \Delta + 2\omega^A\, D_A - \left(\frac{\RR}{2} - \Div \omega - \omega^2\right).
\end{eqnarray}
They are adjoint to each other \cite{eardley-1998-57}: for any section $S$ and complex
functions $f$, $g$ on $S$ we have
\begin{equation}
 \int_S \bar f\, P\,g\,\sqrt{\det q}\, d^2 x = 
 \int_S  g\, P^* \bar f \,\sqrt{\det q}\, d^2 x.
\end{equation}

Given the matter sources (represented by the Einstein tensor), we have a system of 7 differential equations for 12 quantities. All but one
are evolution equations.
Equations (\ref{eqThetaL}) and (\ref{eqThetaK}) appeared in \cite{eardley-1998-57} and
their combination was used in \cite{kavanagh-2006-}

Equation (\ref{eqThetaL})  doesn't involve time derivatives. It is a differential equation for $n_k$ and algebraic for $n_l$. Given a spacetime and a mariginally trapped surface therein, we can use (\ref{eqThetaL}) to construct a family of mariginally trapped
surfaces constituting a horizon. 
 We may solve it by specifying
$n_k$ and then deriving $n_l$ or, if $P$ is invertible, the other way round. 
In both cases we have a freedom of specifying one function.

The character of $H$ is determined by $\tau^\mu\,\tau_\mu = 2n_l n_k$. It is null
if $n_k = 0$ or $n_l = 0$, spacelike if the signs of $n_l$ and $n_k$ are opposite
and timelike if the signs are the same.

If $\sigma^l = 0$ and $G_{\mu\nu} = 0$, we can take $n_k = 0$ and any $n_l$. In this way we obtain a non--expanding horizon. Although we have a complete freedom of specifying one,
positive function $n_l$,  various choices do not lead to different submanifolds of the 
spacetime, but
rather change the slicing of a given $H$. With proper choice of slicing we can make $H$
a weakly isolated horizon. 

If any of those quantities does not vanish and $P$ turns out to be strictly negative, one
can prove that  $H$ is strictly spacelike provided that $n_l > 0$.  This
can be done by applying the
strong maximum principle \cite{andersson-2005-95}. In simple terms, one proves that $P^{-1} f$
is negative if $f > 0$.  Note that although $P$ itself is not invariant with respect
to gauge transformations (\ref{eqGauge}), the property stated above holds even after
rescaling of $l$ and $k$, provided that
we choose $F > 0$ (we don't flip the direction the vectors point).   

In spacelike case different solutions of (\ref{eqThetaL})
lead to truly different horizons, \emph{i.e.} different submanifolds of the spacetime
\cite{ashtekar-2005-}. Furthermore, if $\theta^k < 0$, the hypersurface is a DH.
In \cite{ashtekar-2005-} it is also proved that, under
some reasonable technical assumptions, there are strong restriction concerining the
location of DH's with respect to each other. In particular, they cannot foliate
any neighbourhood of DH.

\subsection{Angular momentum}

Following \cite{ashtekar-2004-7} and \cite{hayward-2006-}, we define the angular momentum
of $H$ with respect to a specified vector field
$X^A$, tangent to a section, as
\begin{equation}
J_X = -\frac{1}{8\pi G}\int_S X^A\,\omega_A\,\sqrt{\det q}\,d^2 x \label{eqJDef}. 
\end{equation}
In the null case this definition agrees completely with the standard one. In the DH case
it is slightly different from the definition in \cite{ashtekar-2004-7}, which in our terms can be written as
\begin{equation}
J'_X = -\frac{1}{8\pi G}\int_S X^A\,\left(\omega_A + \frac{1}{2}D_A \ln\left|\frac{n_l}{n_k}\right|\right)\,\sqrt{\det q}\,d^2 x. 
\end{equation}
However, as was pointed out in \cite{ashtekar-2004-7}, the definition is not reliable if $X^A$ is
not a symmetry, or at least divergence--free. If $D_A X^A = 0$, both definitions actually
coincide. Moreover, (\ref{eqJDef}) is gauge invariant with respect to (\ref{eqGauge}).

\subsection{Small perturbations of an IH}
\label{secSmallPerturbations}

The framework introduced above is particularily well--suited to discuss
the perturbation of an IH. As we mentioned, an IH can be perturbed by matter inflow or
($T_{\alpha\beta} \tau^\alpha \neq 0$) or by external gravitational field ($\sigma_l \neq 0$). In this paper we will analyse the perturbation by a gravitational field in
the linear order. 

External gravitational field, for example a gravitational wave, causes
$\sigma^l \neq 0$. This affects the equations (\ref{eqOmega}) and (\ref{eqQ}), but not (\ref{eqThetaL}), as it involves
 $\sigma^l\,^2$. Therefore in the linear order $n_k$ remains 0.  
 
This has important consequences: first, the second term in (\ref{eqQ}) vanishes and therefore the metric is perturbed only by $n_l \,\sigma_l$. This means that even if the geometry of sections is deformed (by tidal forces), the deformation does not affect their total area.

Moreover, one can integrate equations (\ref{eqQ}) and (\ref{eqOmega}) with the $n_k$ terms absent. 
They yield
\begin{eqnarray}
q_{AB}(t) &=& q_{AB}(t_0)+ \int_{t_0}^{t} \,2n_l(t')\,\sigma^l_{AB}(t')\,dt' + o(\sigma^l)
\label{eqSmallQ} \\
\omega_A(t) &=& \omega_A(t_0)+D^B (q_{AB}(t) - q_{AB}(t_0)) + o(\sigma^l) \label{eqSmallOmega}
\end{eqnarray}
($D^B$ in the second equation is take with respect to $q_{AB}(t_0)$).  
This means that while $\omega$ (and therefore the angular momentum of the horizon)
may change in the result of influence of external gravitational field, its change
is rigidly connected with the change of geometry. If the horizon returns to its original shape after the perturbation, the angular momentum returns to its original value.

The conclusions above are valid for, among other things,  
Booth and Fairhurst's slowly evolving horizons \cite{booth-2004-92, kavanagh-2006-} without matter accretion. Under their assuptions, $\sigma^l$ is of the
 order of $\epsilon \, a_H^{-1/2}$ \cite{booth-2004-92}, $\epsilon$ being the Booth--Fairhurst ``small parameter''
 and $a_H$ denoting the horizon area. However, we do not make any use of the slow
 change of $\theta^k$  $$\left| \Lie_n \theta^k \right| \sim \epsilon \, a_H^{-1}$$ (condition 3 in their papers).

\section{A family of exact solutions}
\label{secExact}

Bartnik and Isenberg have investigated the constraint equations on a DH in the case
of full spherical symmetry \cite{bartnik-2006-23}. We can generalize their result to a special
class of axialy symmetric horizons. 

The set of equations (\ref{eqQ})--(\ref{eqThetaL}) seems very difficult to treat at first glance. However, we may achieve
a great deal of simplification if we assume $n_l$, $n_k$, $\theta_l$, $\theta_k$, $T_{\alpha\beta}\,\tau^\alpha l^\beta$ and $T_{\alpha\beta}\,\tau^\alpha k^\beta$ 
to be constant on sections. In this manner we hope to reduce the equations (\ref{eqQ})--(\ref{eqThetaL}) to
a system of ODE's. For simplicity we also put $\sigma_k = \sigma_l = 0$. This implies, due to 
(\ref{eqQ}), that the horizon geometry
simply scales by a constant conformal factor with time $q(t) = e^{F(t)}\, q_0$. 

The assumption of scalar quantities to be constant on sections imposes further 
conditions on the geometry of $H$. In order to satisfy both (\ref{eqThetaL}) and (\ref{eqThetaK}) we must assume 
$\frac{1}{2}\RR - \Div \omega -\omega^2$ and $\frac{1}{2}\RR + \Div \omega - \omega^2$ to be constant on sections as well.
This cannot be achieved unless 
\begin{equation}
\Div \omega = 0 \label{eqDiv0}
\end{equation}
(on a sphere there are no vector fields of constant and non--vanishing divergence). We therefore have to assume 
\begin{equation}
\frac{1}{2}\RR - \omega^2 = \textrm{const } \label{eqRR}
\end{equation}
on sections. 

We may construct a metric $q_0$ and a smooth vector field $\omega_0$ satisfying (\ref{eqDiv0}) and (\ref{eqRR}) in the following way: assume $q_0$ to be 
axially symmetric
\begin{eqnarray}
q_0 &=& \dd \theta^2 + f(\theta)^2 \, \sin^2\theta\,\dd\varphi^2. \\
f(0)&=&f(\pi) = 1 \nonumber\\
f'(0)&=&f'(\pi) = 0 \nonumber
\end{eqnarray} and $\omega_0$ to be proportional to the axial Killing vector 
$\omega_0 = A(\theta)\, \partial_\varphi$. $\omega_0$ is now automatically divergence--free and it vanishes at the two poles of the metric
$\theta=0$ and $\theta=\pi$. 

In order to solve (\ref{eqRR}) assume that the curvature $\RR$ attains its global minimum $\RRmin$ at the poles, \emph{i.e.} $\RR(\theta) \ge \RRmin$ with equality at $\theta=0$ and $\theta=\pi$. Note that $\RR$ is axially symmetric and therefore always has a local maximum, minimum or is constant near the poles. Moreover, all odd terms of its Taylor expansion at $\theta = 0$ and
$\theta = \pi$ vanish. Therefore
\begin{equation}
A(\theta)=\sqrt{\frac{\RR(\theta) - \RRmin}{2f(\theta)^2\sin^2\theta}},
\end{equation}
is well--defined and smooth everywhere. It also
solves (\ref{eqRR}), because
 $\frac{1}{2}\RR(\theta) - \omega_0\,^2 = \frac{1}{2}\RRmin$.

\begin{figure}
	
	\includegraphics[width=10cm,height=5cm,bb=0 0 583 306]{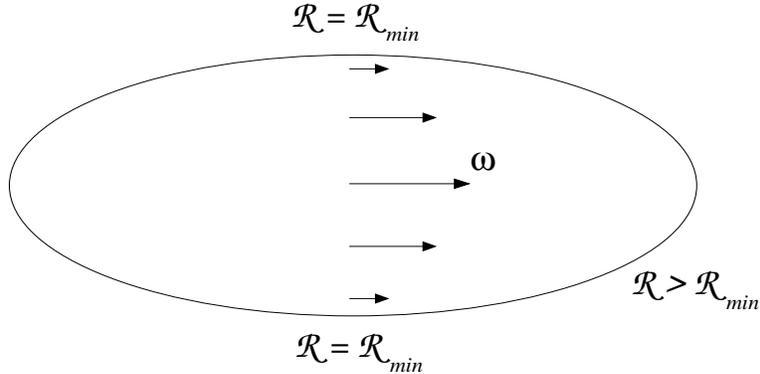}
\caption{Geometry of the horizon section}
\label{figSection}
\end{figure}

The constructed horizon will have, in general, non--vanishing angular momentum with respect to $\partial_\varphi$ (one can construct examples with strictly vanishing $J_\varphi$ as well -- see Appendix \ref{appJ0}). 
It's interesting to note the geometry of sections is necessarily flattened  on
the poles (i.e. $\RR$ is smaller) and rounded in the middle due to its rotation,
just like in the case of a rotating drop of fluid, and in good agreement with intuition (see fig. \ref{figSection}).

We have to deal with the last equation now, (\ref{eqOmega}). We need $\omega$ to rescale together with $q$ in such a way that (\ref{eqRR}) is preserved.
It's straightforward to see that we need the vector $\omega^A(t)$ to
be $e^{-F(t)} \omega_0\,^A$ (or $\omega_A (t) = \omega_0\,_A$ as a one--form).
(\ref{eqOmega}) yields now an equation for $G_{\alpha A} \tau^\alpha$: it must be proportional 
to $\omega$
\begin{equation}
G_{\alpha A}\, \tau^\alpha = \alpha(t) \, \omega_A.
\end{equation}
  
Let's summarize the considerations now. We take the section metric to be $q = e^{2F(t)} q_0$, the one--form $\omega_A = \omega_0\,_A$, where $q_0$ and $\omega_0$ are constant in time and constructed 
according to the prescription stated above. $\RRmin$ is the minimum of the curvature of $q_0$, also constant in time.
We introduce the short--hand notation $G_{\mu\nu}\,\tau^\mu l^\nu = \rho_1(t)$ and $G_{\mu\nu}\,\tau^\mu k^\nu = \rho_2(t)$.
(\ref{eqQ})--(\ref{eqThetaK}) yield the following set of equations:
\begin{eqnarray}
\dot F &=& \frac{1}{2}n_k\,\theta^k \\
\alpha &=& n_k\,\theta^k \\
0 &=& -\frac{\RRmin}{2}\,n_k -\rho_1 e^{2F} \\
\dot \theta^k &=& \rho_2 - \frac{\RRmin}{2}\,n_l\,e^{-2F} - \frac{1}{2} n_k\,\theta^k\,^2.  
\end{eqnarray}
All quantities except $\RRmin$ are functions of time. For a spacelike surface and $\RRmin > 0$ this system of ODE's can be reduced to 
the one in \cite{bartnik-2006-23}, which can be seen by substiting $e^F = \frac{\RRmin}{2}y^{2/3}$, $\rho_1 = -n_k(\rho + \xi)$, $\rho_2 = n_l(\rho - \xi)$,
$n_k\,n_l = -1/2$
and eliminating $\theta^k$.

We may obtain a Cauchy surface of an Einstein metric if we assume $\RRmin$ to be equal to $0$ (sections flat
at the poles). We take $n_k = \rho_1 = \rho_2 = \theta _k = 0$, $F=\const$, $n_l(t)$ and
$n_k(t)$ arbitrary. We can choose them in such a way that the horizon is spacelike.
By comparing with (\ref{eqKAn}) and (\ref{eqKnn}) we conclude that in general the resulting initial data surface is neither time--symmetric, nor extremal. It is not a DH, as expansions of both null
vector fields vanish.

\acknowledgements
The author would like to thank Jerzy Lewandowski, Tomasz Paw\l{}owski, Lars Andersson,
Paul Tod, Piotr Chru\'sciel, Martin Bojowald, Badri Krishnan, Ivan Booth, Jos\'e Luis Jaramillo and others for useful discussions and remarks.

\appendix
\section{Gauge condition for the null ortogonal vectors}
\label{appGauge}

We assume that $k^\mu$ and $l^\mu$ are ortogonal to sections of $H$, which are
preserved by the flow of $n^\alpha$. This is equivalent to 
\begin{eqnarray}
\Lie_n l_\alpha &=& A\, l_\alpha + B\, k_\alpha \\
\Lie_n k_\alpha &=& C\, l_\alpha + D\, k_\alpha \label{eqLieKL}
\end{eqnarray}
for some functions $A$, $B$, $C$ and $D$. Furthermore, the normalization
conditions, differentiated, yield
\begin{eqnarray}
\nabla_n k_\alpha\, k^\alpha &=& \nabla_n l_\alpha\, l^\alpha = 0 \\
\nabla_n k_\alpha\, l^\alpha &=& - \nabla_n l_\alpha \,k^\alpha.  
\end{eqnarray}
It follows that $A = D = 0$ and $B = -C$. Without
loosing generality and in order to simplify the equations we can put $B = 0$. This, substituted
back to (\ref{eqLieKL}), leads directly to (\ref{eqGaugeFixing}). 

\section{Extrinsic curvature of a spacelike surface}
\label{appK}

For a spacelike surface ($n_l > 0$ and $n_k < 0$) we can normalize $\tau$ and $n$
\begin{equation}
\hat \tau^\alpha = (-2n_l\,n_k)^{-1/2}\, \tau^\alpha \qquad \hat n^\alpha = (-2n_l\,n_k)^{-1/2}\, n^\alpha.\
\end{equation}
The extrinsic curvature, \emph{i.e.} the pullback of the convariant derivative 
of $\hat \tau$, expressed in terms of introduced variables, reads:
\begin{eqnarray}
{\cal K}_{AB} &=& n_l L_{AB} + n_k K_{AB} \label{eqKAB}\\
{\cal K}_{\alpha\beta} \hat n^\alpha \hat n^\beta &=& |8n_l n_k|^{-1/2} \Lie_n  \ln\left|\frac{n_l}{n_k}\right| \label{eqKnn} \\
{\cal K}_{A \beta} \hat n^\beta &=& \omega_A + \frac{1}{2}D_A\,\ln\left|\frac{n_l}{n_k}\right|\label{eqKAn}.
\end{eqnarray}

\section{Non--spherically symmetric DH's with vanishing angular momentum}
\label{appJ0}

We aim to construct a non--spherically symmetric dynamical horizon, of the family
presented in Section \ref{secExact}, with strictly vanishing angular momentum. 
This can be achieved if we take
$f(\theta)$ such that 
\begin{enumerate}
\item $\RR(\pi/2 + \theta) = \RR(\pi/2 - \theta)$,   
\item $\RR(\theta) \ge \RRmin$ and $\RR(0)=\RR(\pi/2)=\RR(\pi) = \RRmin$, 
\item all derivatives of $\RR$ vanish at $\theta=\pi/2$,
\end{enumerate}
and take $\omega = A(\theta)\, \partial_\varphi$, with
\begin{equation}
A(\theta) = \pm \sqrt{\frac{\RR(\theta) - \RRmin}{2f(\theta)^2\sin^2\theta}} \label{eqAAA}.
\end{equation}
 The positive sign
is taken for $\theta < \pi/2$ and negative for $\theta \ge \pi/2$. $A(\theta)$ is a smooth function
despite the sign change, due to condition 3.  $J_\varphi$ is 0 due to conditions 1 and 2
and (\ref{eqAAA}).  

Clearly, in this example the matter falling in does have angular momentum but
its inflow is balanced to  0 thanks to different directions of rotation
above and below $\theta=\pi/2$.

\bibliography{ihdh}

\end{document}